\documentclass[review,3p]{elsarticle}

\usepackage{cite}
\usepackage{graphicx}
\usepackage[cmex10]{amsmath}
\usepackage{cases}
\usepackage{array}
\usepackage{amsfonts}
\usepackage{stmaryrd}
\usepackage{url}
\usepackage{wrapfig}
\usepackage{epstopdf} 

\newcommand\prob[1]{P\left(#1\right)}
\newcommand\e{\mbox{e}}
\newcommand\E[1]{\mathbb{E}\left[#1\right]} 
\newtheorem{theorem}{Theorem}
\newtheorem{lemma}{Lemma}
\newproof{pf}{Proof}

\begin{document}

\title{On the Maximal Shortest Path in a Connected Component in V2V}

\author[tsp]{M. Marot\corref{cor}}
\ead{michel.marot@telecom-sudparis.eu}
\cortext[cor]{Corresponding author}

\author[nti]{A.M. Said}
\ead{amounir@nti.sci.eg} 

\author[tsp]{H. Afifi}
\ead{hossam.afifi@telecom-sudparis.eu}

\address[tsp]{Laboratoire SAMOVAR\\ Institut Mines-T\'el\'ecom/T\'el\'ecom SudParis \\
CEA Saclay - Nano Innov, PC 176 - Bât 861\\ 91191 Gif-sur-Yvette Cedex, FRANCE}
\address[nti]{National Telecommunication Institute, Cairo, Egypt}


\begin{abstract}
In this work, a VANET (Vehicular Ad-hoc NETwork) is considered to operate on a simple lane, without infrastructure.
The arrivals of vehicles are assumed to be general with any traffic and speed assumptions. 
The vehicles communicate through the shortest path. In this paper, we study the probability distribution of the number of hops 
on the maximal shortest path in a connected component of vehicles. The general formulation is given for any assumption of road traffic. 
Then, it is applied to calculate the z-transform of this distribution for medium and dense networks in the Poisson case. 
Our model is validated with the Madrid road traces of the Universitat Polit\`ecnica de Catalunya.
These results may be useful for example when evaluating diffusion protocols through the shortest path in a VANET, 
where not only the mean but also the other moments are needed to derive accurate results. 
\end{abstract}


\maketitle

\section{Introduction}
Evaluating Vehicular Ad-hoc NETworks (VANET) protocols is generally complex and difficult as it requires to consider many parameters.
Ideally, a VANET model should take into account at least accurate radio models, accurate road traffic models and protocol models. For this reason, most
VANET protocols are evaluated by simulation. Nevertheless, a formal method to identify shortest hop performance is currently very relevant,
 considering the current requirements and developments for Vehicle-to-Vehicle (V2V) communication in the Intelligent Transportation 
Systems (ITS) domain. Models given connectivity probabilities as a function of the distance between a sender and a receiver avoid complex and long simulations. It is the same for models of the time to 
route the information from a source to a destination or models given the number of hops in the shortest path. For example, it is interesting
to know the probability that a message emitted from a source can reach a vehicle located at a certain distance ahead of the road. If this probability is positive, we
would like to know how much hops are necessary to reach the destination, and how long time is needed. Connectivity probabilities have been already studied (e.g. \citep{altman})
but the process of the number of hops, when following the shortest path, is not exactly known (there are bounds in \citep{soua2012}). The goal of this paper is 
to address this issue.

In this paper, we model the number of hops of the maximal shortest path in a V2V connected component, which is
the shortest path between the vehicle at an end of the connected component to the vehicle at the other extremity.
Modelling the maximal shortest path can be directly used in combination with the connectivity probability to evaluate 
the number of hops in a diffusion protocol for instance. 
Let us call the hop density the average number of hops in a connected component divided by the average 
size of the connected component. Then, assuming a path exists between a sender and a receiver located at a certain distance, 
it can be used to approximate the number of hops between them by multiplying the hop density by the distance separating these two nodes.
Using the results we present in this paper, the hop density can be estimated.
More generally, modelling the maximal shortest path allows to derive the performance of new 
network protocols or mechanisms. Examples of such problems where
models of number of hops in a connected component are useful to evaluate the solution are \citep{adel}, \citep{adel2} and \citep{adel3}). 
In \citep{adel}, we proposed a LTE-assisted D2D solution for dead-ends recovery. The problem is then to calculate the delay to transmit
a message from a sender to the dead-end, the delay of the LTE-assisted solution and the delay between the LTE part and the RSU.

In the next section, works related to VANETs modeling are surveyed, the analytical model formulation is presented in section 3, the explicit calculation in the case of Poisson
road traffic is presented in section 4. Section 5 presents the evaluation and model validation. Section 6 is dedicated to concluding remarks and future works. 
Most of the demonstrations are given in the annexes.

\section{Related works}
Regarding the VANET modeling, when the network is sparse, the information propagation speed may depend on the vehicle speed because the network may work in a delay tolerant fashion. 
In \citep{agarwal} and \citep{wu2009}, models were presented for delay tolerant networks or to investigate the impact of disconnections in vehicle networks 
on the information propagation. These are interesting results but in our present work, we consider only real-time applications. As a consequence, we consider the information
is sent faster than the vehicle speed, so the information transfer is stopped if there is a disconnection.

In works like \citep{veronique} or \citep{coverage}, the number of nodes in a connected component is given. Though these works provide insightful results on cluster sizes,
they did not consider the shortest path. 
The connectivity in VANETs was studied as in \citep{panichpapiboon}, \citep{wu2008}, and \citep{khabazian} where the connectivity probabilities were derived. 
In \citep{miorandi}, authors proposed an analogy 
by which the ad hoc network connectivity is modeled by a G/D/$\infty$ queue. In \citep{altman}, the authors used this analogy to derive the number of cars in a connected component 
of vehicles in a VANET and what they called the connectivity distance, which is the length of a vehicle connected component. In \citep{miorandialtman}, the authors investigated 
the effect of channel randomness on connectivity, with particular emphasis on the effect of lognormal shadowing and Rayleigh fading phenomena. The connectivity analysis 
had been extended to the case of a Nakagami fading channel in \citep{neelakantan2012}. In \citep{neelakantan2013}, authors studied the minimum transmitted power necessary 
to ensure network connectivity in a VANET under BER requirements. 

These papers provide insightful results on connectivity analysis. However, they give no detail on the number of hops in the shortest path of a vehicle connected component, 
which is important to estimate propagation delays. In \citep{abdrabou}, the multi-hop packet delivery was approximated through an effective bandwidth approach, with the 
traffic parameters approximated by average ON and OFF periods. In \citep{zhang2012} and \citep{soua2012}, the connectivity and the number of hops through the shortest path 
were exactly modeled for Poisson traffics. Unfortunately, the results were given as a recursive formula, which makes it difficult to perform the analysis as a
function of network parameters. 
In this paper, we provide a method to calculate the number of hops through the shortest path in a connected component in VANETs for any traffic assumption. We validate it
against VANET traces provided byt the Universitat Polit\`ecnica de Catalunya. We illustrate this method by deriving the explicit expression of the z-transform of its 
probability in the case if Poisson traffic assumption. In this case, we provide a closed-form  expression which is valid for medium and dense networks (i.e. for networks 
for which the product of the node density by the coverage radius is larger than $\ln 4$). This model can then be used to derive access delays, or path lengths.

\section{Analytical model}
\subsection{Model and assumptions}

In this paper, we study the number of hops in a connected component. At a given time, vehicles are spread on the road, separated by random distances. Considering
a solid disk propagation model, which may be refined later, vehicles can communicate if the interdistance between them is less than a certain coverage radius. In this
case, vehicles can constitute connected components. The goal is to model the distribution of the number of hops of the maximal shortest path in a connected component, which is
the shortest path between the vehicle at an end of the connected component to the vehicle at the other extremity. Note that the maximal shortest path is not the number
of hosts located between the sender and the receiver at both extremities because of the shortest path routing algorithm. Actually, we assume a routing protocol is used allowing only
vehicles on the shortest path to retransmit packets. In other words, if two vehicles receive at the same time the same packet from a sender, only the furthest one under the
sender's coverage forwards the packet.

We consider a vehicles traffic on a straight road.
If the vehicle transmission range is sufficiently large compared to the road width, this commonly used assumption makes sense.
We do not make any assumption on the arrival process nor the speeds. At a given time $t$, the vehicles are separated by a distance $\{I_k\}_{k>0}$.
The process $\{I_k\}_{k>0}$ is considered independent and identically distributed. Its cumulative distribution function is $P(I_k\leq x)=F(x)$ and 
its density $f(x)=\frac{dF(x)}{dx}$. Note that while the interdistances are considered independent, the traffic it models is not independent. It is so only when $F$ corresponds
to the exponential distribution. The vehicles arrive at position 0 of the road. 

As in \citep{altman}, to study the connectivity, we use the equivalent infinite server queueing model presented by Moriandi and Altman in \citep{miorandi}.
These authors proved that the problem of calculating the number of wireless nodes in a connected cluster and the length of the connected path at a given 
time with the solid disc transmission range model is equivalent to the problem of calculating the number of clients
in a busy period of the G/D/$\infty$ queue and the duration of this busy period: the busy period of the G/D/$\infty$ queue 
corresponds to a connected vehicle cluster, the time is replaced by the distance, and the communication range replaces the service time.

By using the results given by Liu and Shi in \citep{liushi} on the probability generating function of the number of clients in a busy period and on
the Laplace transform of the busy period duration, Yousefi et al. characterize in \citep{altman} the number of clients in a vehicle connected component
and the connectivity distance, that is the length of a vehicle connected component.
Here, we adopt the same approach modeling the connectivity in terms of the GI\textsuperscript{x}/D/$\infty$. Nevertheless, since we consider the number of hops 
on the shortest path and not each vehicle in the path and second because we do not make any assumption on the vehicle traffic, the input process of the queue
is no more the Poisson process described in \citep{altman} but a sampled process of this Poisson process. As we will see in the next section, 
it is a Markovian process.

\subsection{The interdistance process}
\label{paragrapheinterdistanceprocess}

\begin{figure}[!t]
\centering
\includegraphics[width=3.5in]{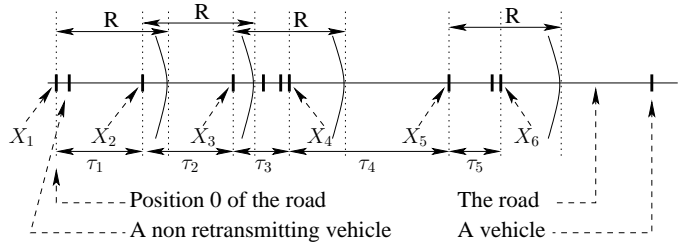}
\caption{Example showing the sampling of the vehicles according to the shortest path}
\label{dessinvoitures}
\end{figure}

Let $X_1$, $X_2,\hdots,X_n,\hdots$ be the {\bf retransmitting} vehicles. $X_{n+1}$ is the farthest vehicle under $X_n$'s coverage, or the next node after 
$X_n$ if the distance between $X_n$ and $X_{n+1}$ is larger than $R$.
Because of the shortest path routing algorithm, 
if $X_{n+1}$ is the farthest vehicle under $X_n$ coverage, 
it is assumed to retransmit $X_n$ packets even if there are other vehicles between them.
By extension, $X_n$ denotes also the coordinate (location) of $X_n$ on the road.
Let $\tau_n$ be the distance between node $X_n$ and node $X_{n+1}$.
Let $I_{n,k}$ the $k^{\mbox{th}}$ interdistance after the $n^{th}$ vehicle.
First of all, we need to characterize the process of the distance $\tau_n$ separating to retransmitting nodes.

\begin{theorem}
\label{interarrivees}
When $R-x_{n-1}\leq x_n\leq R$, the cumulative distribution function of $\tau_n$ knowing $\tau_{n-1}=x_{n-1}$ and $\tau_{n-1}\leq R$ 
is given by:
\begin{eqnarray}
\label{eqgenericinterdist1}
\lefteqn{F_{\tau_n/\tau_{n-1}}(x_n,x_{n-1})}\nonumber\\
& = &\frac{1}{1-F(R-x_{n-1})}\displaystyle\int_{u=R-x_{n-1}}^{x_n}{\cal L}^{-1}\left[\frac{{\cal L}\left[f(x+R-x_{n-1})\right]}{1-{\cal L}\left[f(x)\right]}\right]
_{\left(u-(R-x_{n-1})\right)}\left[1-F(R-u)\right]du\\
\label{eqgenericinterdist1a}
\lefteqn{F_{\tau_1}(x_1)}\nonumber\\
& = &\displaystyle\int_{u=0}^{x_n}{\cal L}^{-1}\left[\frac{{\cal L}\left[f(x)\right]}{1-{\cal L}\left[f(x)\right]}\right]_{(u)}\left[1-F(R-u)\right]du
\end{eqnarray}
where ${\cal L}$ and ${\cal L}^{-1}$ denotes the Laplace transform and its inverse.

When $x_n\geq R$, the cumulative distribution function of $\tau_n$ knowing $\tau_{n-1}=x_{n-1}$ and $\tau_{n-1}\leq R$ 
is given by:
\begin{eqnarray}
\label{eqgenericinterdist1bis}
F_{\tau_n/\tau_{n-1}}(x_n,x_{n-1}) & = & P\left(I_{n,1}\leq x_n/I_{n,1}>R-x_{n-1}\right)\nonumber\\
& = & \frac{F(x_n)-F(R-x_{n-1})}{1-F(R-x_{n-1})}\\
\label{eqgenericinterdist1bisa}
F_{\tau_1}(x_1) & = & F(x_1)
\end{eqnarray}

\end{theorem}

\begin{pf}
See Appendix~\ref{demonstrationthminterarrivees}.
\end{pf}

We can observe from equations (\ref{eqgenericinterdist1}), (\ref{eqgenericinterdist1a}), (\ref{eqgenericinterdist1bis}) and (\ref{eqgenericinterdist1bisa}) that the distance between the 
retransmitting vehicles on the shortest path is actually a markovian process.
In the following sections of the paper, we assume, without loss of generality, $X_1$ at the beginning of a connected component, which
mean that the node $X_0$ before $X_1$ on the x-axis is out of range of $x_1$: $X_1-X_0>R$.

In the special case of a Poisson process with rate $\lambda$, 
\begin{eqnarray}
\label{FPoisson}
F(x) & = & 1-e^{-\lambda x}
\end{eqnarray}

and thus,

\begin{eqnarray}
\label{eqinterar1}
\forall x_1\leq R, F_{\tau_1}(x_1) & = & \e^{-\lambda(R-x_1)}\left(1-\e^{-\lambda x_1}\right)\\
\label{eqinterar2}
\forall x_1\geq R, F_{\tau_1}(x_1) & = & 1-\e^{-\lambda x_1}
\end{eqnarray}

Also,
\begin{eqnarray}
\label{eqinterar3}
\forall x_n;R-x_{n-1}\leq x_n\leq R, F_{\tau_n/\tau_{n-1}}(x_n,x_{n-1}) & = & \e^{-\lambda(R-x_n)}-\e^{-\lambda x_{n-1}}\\
\label{eqinterar4}
\forall x_n\geq R, F_{\tau_n/\tau_{n-1}}(x_n,x_{n-1}) & = & 1-\e^{-\lambda\left[x_n-(R-x_{n-1})\right]}
\end{eqnarray}

In the special case of a mixture of two exponential variables with rates $\lambda_1$ and $\lambda_2$ and weights $\alpha_1$ and $\alpha_2=1-\alpha_1$,
\begin{eqnarray}
\label{FMixtudeExponetials}
F(x) & = & 1-\alpha_1 e^{-\lambda_1 x}-\alpha_2 e^{-\lambda_2 x}
\end{eqnarray}

and thus, by denoting $\lambda_m=\lambda1(1-\alpha_1)+\lambda_2(1-\alpha_2)$,

\begin{eqnarray}
\forall x_1\leq R, F_{\tau_1}(x_1) & = & \frac{\lambda_1 \lambda_2}{\lambda_m}\left[\alpha_1\frac{e^{-\lambda_1 R}}{\lambda_1}\left(e^{\lambda_1 x}-1\right)
+\alpha_2\frac{e^{-\lambda_2 R}}{\lambda_2}\left(e^{\lambda_2 x}-1\right)\right] \nonumber\\
& & + \left[\alpha_1\lambda_1\left(1-\frac{\lambda_2}{\lambda_m}\right)+\alpha_2\lambda_2\left(1-\frac{\lambda_1}{\lambda_m}\right)\right]\nonumber\\
\label{eqinterar1ExpoMixture}
& & \times \left[\frac{\alpha_1 e^{-\lambda_1 R}}{\lambda_m-\lambda_1}\left(1-e^{-(\lambda_m-\lambda_1)x_1}\right)
+\frac{\alpha_2 e^{-\lambda_2 R}}{\lambda_m-\lambda_2}\left(1-e^{-(\lambda_m-\lambda_2)x_2}\right)\right]\\
\forall x_1\geq R, F_{\tau_1}(x_1) & = & 1-\alpha_1 e^{-\lambda_1 x_1}-\alpha_2 e^{-\lambda_2 x_1}
\label{eqinterar2ExpoMixture}
\end{eqnarray}

Also,
\begin{eqnarray}
\label{eqinterar3ExpoMixture}
\forall x_n;R-x_{n-1}\leq x_n \leq R, & &\nonumber\\
 F_{\tau_n/\tau_{n-1}}(x_n,x_{n-1}) & = &\alpha_1\frac{\lambda_2}{\lambda_m}\left[e^{-\lambda_1(R-x_n)}-e^{-\lambda_1x_{n-1}}\right]+
\alpha_2\frac{\lambda_1}{\lambda_m}\left[e^{-\lambda_2(R-x_n)}-e^{-\lambda_2x_{n-1}}\right]\nonumber\\
& & +\frac{\alpha_1\lambda_1\left(1-\frac{\lambda_2}{\lambda_m}\right)e^{-\lambda_1(R-x_{n-1})}+\alpha_2\lambda_2\left(1-\frac{\lambda_1}{\lambda_m}\right)e^{-\lambda_2(R-x_{n-1})}}
{\alpha_1 e^{-\lambda_1 (R-x_{n-1})}+\alpha_2 e^{-\lambda_2 (R-x_{n-1})}}\nonumber\\
& & \times\left[\frac{\alpha_1e^{-\lambda_1R}}{\lambda_m-\lambda_1}\left(e^{-(\lambda_m-\lambda_1)(R-x_{n-1})}-e^{-(\lambda_m-\lambda_1)x}\right)\right.\nonumber\\
& &\left.+\frac{\alpha_2e^{-\lambda_2R}}{\lambda_m-\lambda_2}\left(e^{-(\lambda_m-\lambda_2)(R-x_{n-1})}-e^{-(\lambda_m-\lambda_2)x}\right)\right]e^{\lambda_m(R-x_{n-1})}\\
\label{eqinterar4ExpoMixture}
\forall x_n\geq R, F_{\tau_n/\tau_{n-1}}(x_n,x_{n-1}) & = & 1-\frac{\alpha_1 e^{-\lambda_1 x_n}+\alpha_2 e^{-\lambda_2 x_n}}{\alpha_1 e^{-\lambda_1 (R-x_{n-1})}+\alpha_2 e^{-\lambda_2 (R-x_{n-1})}}
\end{eqnarray}

\subsection{Recall on the results of Liu and Shi in \citep{liushi} used in \citep{altman}}

Liu and Shi consider a GI\textsuperscript{x}/G/$\infty$ queue whose customers arrive in batches with generating function $A(z)$, separated by interarrival times $\tau_n$
with distribution function $F(x)$ and service time distribution $H(x)$. $Y_n$ is the number of clients in the $n$\textsuperscript{th} batch and
$T_n$ is the time necessary to complete the services of the $n$\textsuperscript{th} batch. $K(x)=A\left(H(x)\right)$ is the distribution function of $T_n$.
In this case, they show that the z-transform of the distribution function of the number $N_b$ of clients in a connected component is 
\begin{eqnarray}
\label{liushieq1}
\sum_{k=1}^{+\infty}\prob{N_b=k}z^k
& =&  \sum_{n=1}^\infty\left[A(z)\right]^n \int_0^\infty\hdots\int_0^\infty\prod_{i=1}^{n-1}\displaystyle\left\{
\left[K\left(\sum_{j=i}^nx_i\right)
-K(x_i)\right]dF(x_i)\right\}K(x_n)dF(x_n)\quad
\end{eqnarray}

This is true because the probability distribution of the number of customers served in a busy period is
\begin{eqnarray}
\label{liushieq2}
P\left(N_b=k\right)& = & \sum_{n=1}^\infty \prob{\sum_{i=1}^nY_i=k}\times
\prob{\tau_i\leq T_i<\sum_{j=i}^n\tau_j,(i=1,\hdots,n-1),\tau_n>T_n}\quad
\end{eqnarray}

Actually, taking the generating function of (\ref{liushieq2}) and noting the discrete convolutions involved in the generating function, (\ref{liushieq1})
can be obtained.

Also, in the special case of the GI\textsuperscript{x}/D/$\infty$, $K(x)=1$ if $x\geq R$ and 0 otherwise so that (\ref{liushieq1}) becomes
\begin{eqnarray}
\label{liushieq3}
\sum_{k=1}^{+\infty}\prob{N_b=k}z^k
& =&  \sum_{n=1}^\infty\left[A(z)\right]^n \int_{x_1=0}^R\hdots\int_{x_{n-1}=0}^R\int_{x_n=R}^\infty\prod_{i=1}^ndF(x_i)
\end{eqnarray}

and thus Liu and Shi prove that formula (\ref{liushieq1}) simplifies  into:
\begin{eqnarray}
\label{liushieq4}
\sum_{k=1}^{+\infty}\prob{N_b=k}z^k & = & z.\frac{1-F(R)}{1-zF(R)}
\end{eqnarray}

With the same kind of arguments, Liu and Shi present the Laplace transform of the busy period for the GI\textsuperscript{x}/G/$\infty$ which reduces, in
the special case of the GI\textsuperscript{x}/D/$\infty$, to
\begin{eqnarray}
\label{liushieq5}
{\cal L}\left(f_d\right)(s)
&=& \frac{e^{-sR}\left(1-F(R)\right)}{1-\int_0^Re^{-sx}dF(x)}
\end{eqnarray}

The authors of \citep{altman} rely on (\ref{liushieq4}) and (\ref{liushieq5}) to derive an upper bound for the tail distribution of the number of hops
in the connected component:
\begin{eqnarray}
\prob{N_b\geq h} & \leq & \left(1-e^{-\lambda R}\right)^h
\end{eqnarray}
and the average distance of a vehicles connected component $L_{cc}$, which they call connectivity distance of vehicles and which is equal to the average 
length of the busy period:
\begin{eqnarray}
\E{L_{cc}} & = & \frac{1-e^{-\lambda R}}{\lambda e^{-\lambda R}}.
\end{eqnarray}

\subsection{Number of hops in the maximal shortest path}
Due to the independance of the interarrivals of the GI\textsuperscript{x}/G/$\infty$, the multiple integral in (\ref{liushieq3}) can be easily
simplified to obtain (\ref{liushieq4}). However, in our case, considering the shortest path leads to a markovian process as already explained in 
\S.\ref{paragrapheinterdistanceprocess}.

The process of the interdistance between the vehicles is thus not an independent process but a markovian one. 
As a consequence, to compute the distribution of the number of vehicles in a connected component, the results of \citep{altman} cannot be used.
The authors of \citep{altman} rely on the results in \citep{liushi} for the GI\textsuperscript{x}/G/$\infty$, where the arrivals 
are assumed independent. Consequently, to compute the distribution of the number of vehicles in a connected component, the results given in \citep{altman} 
is not compatible with the new process description. That is why we roll back to the approach initiated in \citep{altman} to adapt the model according to the Markovian arrival process.
Therefore, the results presented in \citep{liushi} cannot be directly applied but they must be adapted to the non independent case. 

\begin{theorem}
In the case of a markovian interdistance process, and still denoting by $F_{\tau_i/\tau_{i-1}}(x_i,x_{i-1})$ the cumulative distribution function of
$\tau_i$ knowing $\tau_{i-1}$, the probability to have $k$ hops in a connected component is:
\begin{eqnarray}
\label{distributionDuNombreDeSauts}
\lefteqn{\prob{N_b=k}}\\
& = & \int_{x_1=0}^R\int_{x_2=R-x_1}^R\hdots\int_{x_{k-1}=R-x_{k-2}}^R \left(1-
F_{\tau_k/\tau_{k-1}}(R,x_{k-1})\right)
dF_{\tau_1}(x_1)\prod_{i=2}^{k-1}dF_{\tau_i/\tau_{i-1}}(x_i,x_{i-1})\nonumber
\end{eqnarray}
\end{theorem}

\begin{pf}
By following the same approach as in \citep{liushi}, but using the Markov property, we obtain:
\begin{eqnarray}
\prob{N_b=k}
& = & \prob{\forall i\in \llbracket 1;k-1\rrbracket,\tau_i\leq R<\sum_{j=i}^k\tau_j,\tau_k>R}\\
& = &\prob{\forall i\in \llbracket 1;k-1\rrbracket,\tau_i\leq R,\tau_k>R}\nonumber\\
& = &\int_{x_1=0}^R\prob{\forall i\in \llbracket 2;k-1\rrbracket,\tau_i\leq R,\tau_k>R/\tau_1=x_1}
d\prob{\tau_1=x_1}\nonumber\\
&   & \vdots\nonumber\\
& = &\int_{x_1=0}^R\int_{x_2=R-x_1}^R\hdots\int_{x_{k-1}=R-x_{k-2}}^R
\prob{\tau_k>R/\tau_{k-1}=x_{k-1}}\times\nonumber\\
& &\qquad\qquad\prod_{i=2}^{k-1}d\prob{\tau_i=x_i/\tau_{i-1}=x_{i-1}}d\prob{\tau_1=x_1}\nonumber
\end{eqnarray}
\end{pf}

\section{Explicit calculation in the case of Poisson road traffic}

In this section, we derive the expression of the probabilities $P(N_b=k)$ in the case where the road traffic is Poisson with parameter $\lambda$. 
The Poisson assumptions means that the vehicles are independent, which is usually not the case. 
Nevertheless, the arrival times are often independent, even if the behavior of the cars become not independent anymore when they 
 are on the same road. Moreover, in free flow regime, when the arrivals are separated by time arrivals larger than 5 or 6 seconds, 
they may be independent. It is the case in highways for example and this is why the Poisson assumption is often considered in performance
evaluations. For these situations, we give the explicit solution of the model.
Let us denote $\lambda'=\lambda R$, $\rho=\lambda'e^{-\lambda'}$ and $\rho'=e^{-\lambda'}$.\\

In the special case of the interdistance process described in \S.\ref{paragrapheinterdistanceprocess} (i.e. equations 
(\ref{eqinterar1}), (\ref{eqinterar2}), (\ref{eqinterar3}) and (\ref{eqinterar4})), the probability of the number of hops in the
connected component becomes with a simple change of variable:
\begin{eqnarray}
\prob{N_b=k} &=& \rho^{k-1}
\int_{u_1=0}^1\int_{u_2=1-u_1}^1\hdots\int_{u_{k-1}=1-u_{k-2}}^1\prod_{i=1}^{k-2}e^{\lambda'u_i}\prod_{i=1}^{k-1}du_i\quad
\end{eqnarray}

In order to calculate this integral, it is convenient to denote

\begin{eqnarray}
\mathfrak{M}_{\alpha,k} &=& \rho^k
\int_{u_1=0}^1\int_{u_2=1-u_1}^1\hdots\int_{u_k=1-u_{k-1}}^1u_k^\alpha e^{\lambda'u_k}
\prod_{i=1}^{k-1}e^{\lambda'u_i}\prod_{i=1}^{k}du_i
\end{eqnarray}

All the probabilities $\prob{N_b}$ are function of such integrals. We can observe that 
\begin{eqnarray}
\label{probNbk}\prob{N_b=k} &=
&\rho\mathfrak{M}_{1,k-2}.
\end{eqnarray}

These probabilities can be calculated by reccurence. Actually, by integrating by parts, 
\begin{numcases}{}
 \label{M0k}\mathfrak{M}_{0,k} =\mathfrak{M}_{0,k-1} -\rho \mathfrak{M}_{1,k-2}\\
 \label{Malphak}\mathfrak{M}_{\alpha,k} = \mathfrak{M}_{0,k-1}-\frac{\alpha}{\lambda'}\mathfrak{M}_{\alpha-1,k}-\frac{\rho\mathfrak{M}_{\alpha+1,k-2}}{\alpha+1}
\end{numcases}

Now, for medium and dense networks, that is $\lambda R\geq \ln 4$, a closed form expression can be given for the z-transform of these probabilities.
Let $Q(z)$ be the z-transform of $N_b$ and $M_1(z)$ the z-transform of $\mathfrak{M}_{1,k}$:
\begin{eqnarray}
Q(z) & = & \sum_{k=1}^{+\infty}\prob{N_b=k}z^k\\
M_1(z) & = & \sum_{k=1}^{+\infty}\mathfrak{M}_{1,k}z^k
\end{eqnarray}

\begin{theorem}
\label{casspecial}
In the special case where $\lambda R\geq \ln 4$, the z transform $M_1(z)$ can be expressed in a close formula:
\begin{eqnarray}
M_1(z)
\label{M1dezCasSpecial} & = & \frac{h_1(z)+h_2(z)+h_3(z)}{\rho z^2\left[1+\sqrt{1-4\rho' z^2}-2ze^{\frac{1}{2}\lambda'\left(\sqrt{1-4\rho' z^2}-1\right)}\right]}
\end{eqnarray}
where
\begin{eqnarray}
h_1(z) & = & \sqrt{1-4\rho'z^2}\left[\left(1-\rho'-\rho\right)z^3-\left(1-\rho'\right)z^2
-z\left(1-\rho\right)+2-\rho'-\rho\right]\nonumber\\
h_2(z) & = & e^{\frac{1}{2}\lambda'\left(\sqrt{1-4\rho' z^2}-1\right)}\left[2\rho z^3+2\rho' z^2-z-1
+(z-1)\sqrt{1-4\rho' z^2}\right]\nonumber\\
h_3(z) & = & z^3(\rho'+\rho-1)+z^2(1-3\rho'-2\rho)
+z(1-\rho)+\rho'+\rho\nonumber
\end{eqnarray}
\end{theorem}

\begin{pf}
See \ref{demonstrationGrandtheoremeCasSpecial}.
\end{pf}

The moments of the number of hops can simply be obtained by differentiating Q(z).
\begin{theorem}
The z-transform of $N_b$ is:
\begin{eqnarray}
Q(z) & = & \rho'z+\rho z^2\left(1+M_1(z)\right).
\end{eqnarray}
\end{theorem}


\begin{pf}
\begin{eqnarray}
P\left(N_b=1\right) & = & \prob{\tau_1>R}\\
&=&e^{-\lambda R}\nonumber
\end{eqnarray}
Moreover,
\begin{eqnarray}
\prob{N_b=2}
 & = & \int_{x_1=0}^R\prob{\tau_2>R/\tau_1=x_1}d\prob{\tau_1=x_1}\\
&=&\lambda Re^{-\lambda R}\nonumber
\end{eqnarray}

Then, by using (\ref{probNbk}), we get the result.
\end{pf}

For example, the expectation of $N_b$ is 
\begin{eqnarray}
\label{moyenneNb}
\E{N_b} & = & \displaystyle\frac{e^{+\frac{\lambda'}{2}\sqrt{1-4\rho'}}-\frac{2}{1+\sqrt{1-4\rho'}}\displaystyle e^{-\frac{\lambda'}{2}}}
{e^{\frac{\lambda'}{2}}-\frac{2}{1+\sqrt{1-4\rho'}}\displaystyle e^{+\frac{\lambda'}{2}\sqrt{1-4\rho'}}}
\end{eqnarray}

\section{Simulations and validations}

In order to validate the proposed model, we ran a number of simulations\footnote{The corresponding code is available at http://www-public.it-sudparis.eu/~marot/archiveForReviewers.tar.gz} and compared these results to the analytical results. 
We also compared the model with traces available on the WEB
and provided by the Universitat Polit\`ecnica de Calaunya. 

\subsection{Comparison between simulations and the model}

We compare the model and the simulations in two different cases. First of all, we consider the hyperexponential case where the distance between the vehicles follows a mixture of 
two exponential distributions. It corresponds to the distribution (\ref{FMixtudeExponetials}). Then, we focus on the Poisson case.

\subsubsection{The hyperexponential case}

\begin{figure}[!t]
\centering
\resizebox{5in}{!}{\input{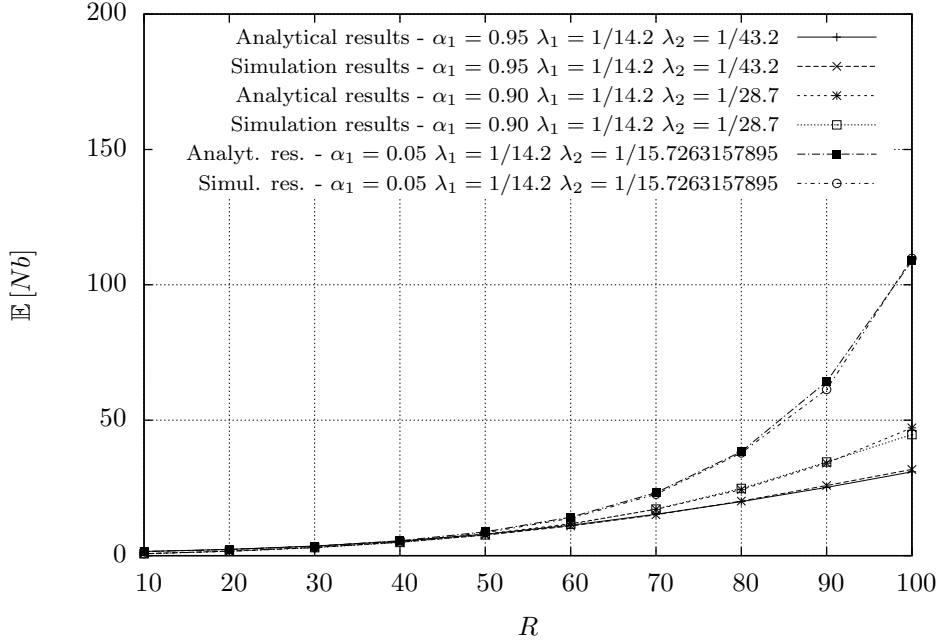}}
\caption{Average number of hops in a connected component, in function of the coverage radius $R$, for different values of parameters $\alpha_1$, $\lambda_1$ and $\lambda_2$ (with $\alpha_2=1-\alpha_1$) at constant mean rate}
\label{comparisonwithsimulations}
\end{figure}

The fig.~\ref{comparisonwithsimulations} presents the comparison of the model and the simulations for the hyperexponential case. Knowing the distribution of the 
distance between the vehicles given by (\ref{FMixtudeExponetials}), we can determine the distribution of the distance between two hops in a connected component
by using (\ref{eqgenericinterdist1}), (\ref{eqgenericinterdist1a}), (\ref{eqgenericinterdist1bis}) and (\ref{eqgenericinterdist1bisa}). The result is given in 
(\ref{eqinterar1ExpoMixture}), (\ref{eqinterar2ExpoMixture}), (\ref{eqinterar3ExpoMixture}) and (\ref{eqinterar4ExpoMixture}). Then, we can evaluate 
(\ref{distributionDuNombreDeSauts}) with a Monte-Carlo method. This curve is presented on the fig.~\ref{comparisonwithsimulations} for different values of 
$\alpha_1$, $\lambda_1$ and $\lambda_2$ (with $\alpha_2=1-\alpha_1$). Note that the average of the three distributions is the same, that is the mean rate of the traffic remains the
same while the parameters of the exponential mixtures are varied. As expected, the size of the connected component decreases when the traffic becomes more and more
bursty. Conversely, it increases with $R$. The simulations and the model fit very well.  

\begin{figure}
\centering
\resizebox{5in}{!}{\input{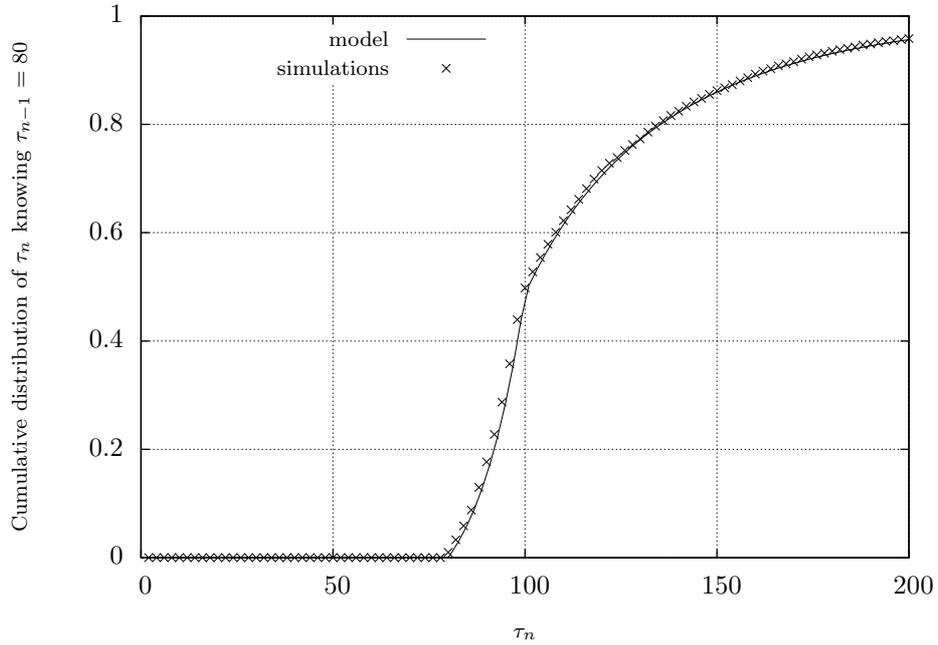}}
\caption{Distribution of $\tau_n$ knowing $\tau_{n-1}=80$ for $R=100$, $\alpha_1=0.95$, $\lambda_1=1/14.2$ and $\lambda_2=1/43.2$}
\label{interdistanceBetweenHopsKnowingTauminus1}
\end{figure}

The validity of (\ref{eqinterar3ExpoMixture}) and (\ref{eqinterar4ExpoMixture}) is presented Fig.~\ref{interdistanceBetweenHopsKnowingTauminus1}. 
It represents the distribution of the distance between two hops $\tau_n$ knowing $\tau_{n-1}$, for the analytical model and for a simulation 
with $\tau_{n-1}=80$, $R=100$, $\alpha_1=0.95$, $\lambda_1=1/14.2$ and $\lambda_2=1/43.2$.

\subsubsection{The Poisson case}
\begin{figure}[!t]
\centering
\resizebox{5in}{!}{\input{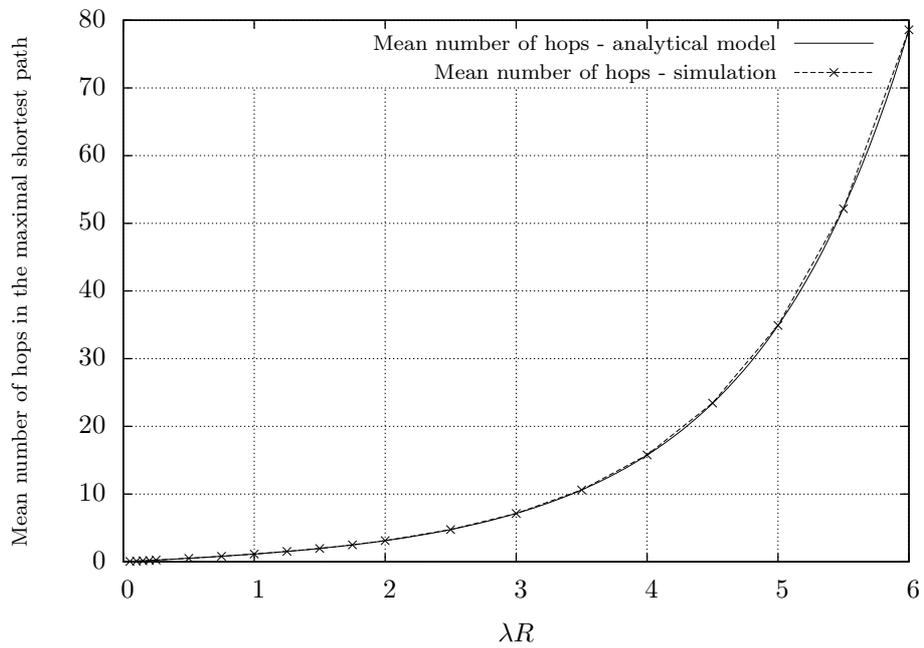}}
\caption{Average number of hops in a connected component, 1st vehicle excluded, in function of $\lambda'$}
\label{meannumberofhops}
\end{figure}

	We compare the results of the analytical model, and more precisely the average number of hops in a connected component given by (\ref{moyenneNb}), with the simulations. 
Figure~\ref{meannumberofhops} shows the average number of hops in a connected component, minus the first node,
in function of different values for $\lambda'$. There is a perfect match between the model and the simulations.
It is interesting to note that the expression of the average number of hops given in equation (\ref{moyenneNb}), though theoretically valid only for $\lambda'>\ln 4$, 
seems to be still valid for any $\lambda'$.
Actually, we can notice that for $\lambda'\in[0:\ln 4]$ when injecting $i\sqrt{4e^{-\lambda'}-1}$ (where, here, $i^2=-1$) in place of $\sqrt{1-4e^{-\lambda'}}$ in eq.~(\ref{moyenneNb}),
the imaginary part disapears so that, it becomes:
\begin{eqnarray}
\label{moyenneNblambdapetit}
\displaystyle\frac{e^{+\frac{\lambda'}{2}\sqrt{1-4\rho'}}-\frac{2}{1+\sqrt{1-4\rho'}}\displaystyle e^{-\frac{\lambda'}{2}}}
{e^{\frac{\lambda'}{2}}-\frac{2}{1+\sqrt{1-4\rho'}}\displaystyle e^{+\frac{\lambda'}{2}\sqrt{1-4\rho'}}} 
&= & \displaystyle\frac{2e^{-\frac{\lambda'}{2}}\cos\frac{\lambda'\sqrt{4e^{-\lambda'}-1}}{2}-1}{2-e^{\frac{\lambda'}{2}}\left[\cos\frac{
\lambda'\sqrt{4e^{-\lambda'}-1}}{2}+\sqrt{4e^{-\lambda'}-1}\sin\frac{\lambda'\sqrt{4e^{-\lambda'}-1}}{2}\right]}
\end{eqnarray}

\subsection{Comparison with traces of the Universitat Polit\`ecnica de Catalunya}
To validate our results, we compared our model to vehicular traces provided by the Universitat Polit\`ecnica de Catalunya 
(cf. \citep{marcofiore} and \citep{marcofiorelink}).
Since this traffic is not Poisson, we determined the distribution of the distance between vehicles and then we injected this distance in (\ref{eqgenericinterdist1}),
(\ref{eqgenericinterdist1a}), (\ref{eqgenericinterdist1bis}) and (\ref{eqgenericinterdist1bisa}) in order to determine the distribution of the distance between hops.
Using this estimated distribution, we calculated the mean of the number of hops in the maximal shortest path by numerically evaluating (\ref{distributionDuNombreDeSauts}) 
with a Monte-Carlo method. With the distribution of the distances between vehicles we just estimated, 
we also generated artificially the same traffic and finally we compared the size of the 
number of hops in the maximal shortest path for the traces, for the artificially generated traffic and the model (i.e. (\ref{distributionDuNombreDeSauts})).

\subsubsection{Traces analysis}
As explained in \citep{marcofiore}, these traces are derived from high-detail real-world traffic counts and describe the road traffic on two highways around Madrid, Spain, 
at several hours of different working days. The Autov\'{\i}a A6 is a motorway that connects the city of A Coru\~na to the city of
Madrid. This road enters into the urban area from the northwest and collects the traffic demand of the conurbation that was built along it. The data collection point is placed
around the 11-km milepost (Madrid direction), where the A6 features three lanes with a speed limit of 120 km/h. However, this location is right after a popular entrance 
ramp that joins the rightmost lane and significantly slows down the road traffic. The length of the considered road segment is 10km.

The road traffic data set is formed by mobility traces for 4 dates and 2 time intervals (8h am and 11h am).  To avoid transient effects, we did not consider the beginning
of the files but we kept the data only after 600 seconds of measurements. Then, since the four data sets are more or less homogeneous at a given
time interval (i.e. either 8h am or 11h am), we concatenated the remaining parts of the corresponding four data sets. Also, we sampled the state of the traffic every 600s.

We approximated the empirical cumulative distributions function $F_{8h}(x)$ and $F_{11h}(x)$ of the distance between each vehicle in each sample (8h or 11h) by the following 
hyperexponential distributions:
\begin{eqnarray}
\label{F8h}
F_{8h}(x) & = & 1-0.92 e^{-x/12.1534}-0.08 e^{-x/40.4655}\\
\label{F11h}
F_{11h}(x) & = & 1 - 0.8100133 e^{-x/16.20606}-0.1899867 e^{-x/44.4476}
\end{eqnarray}

\begin{figure}[!t]
\centering
\resizebox{5in}{!}{\input{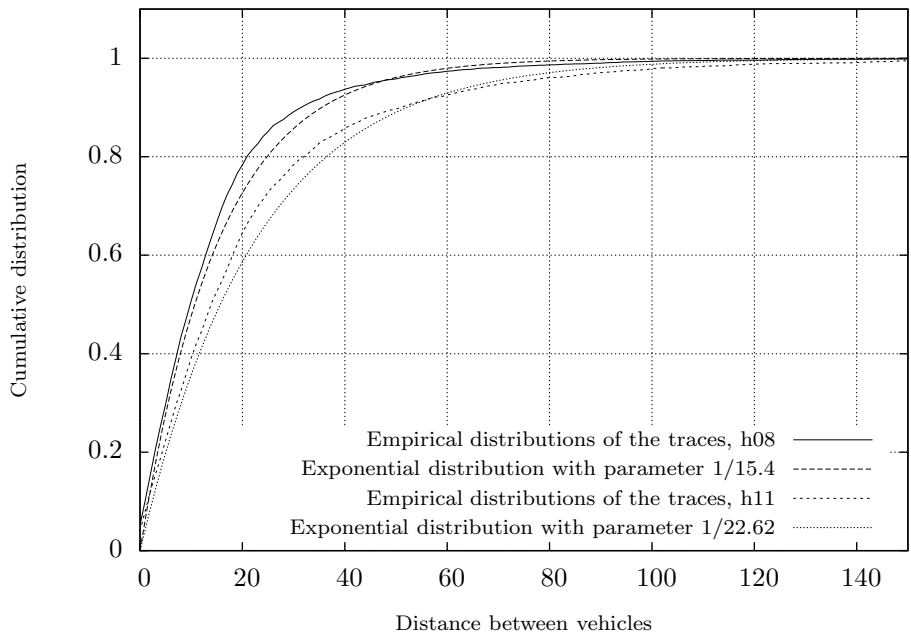}}
\caption{Cumulative distribution function of the distance (in meters) between the vehicles for the traces and for the corresponding exponential distributions with same means}
\label{tracesexponentialinterdistances}
\end{figure}

\begin{figure}[!t]
\centering
\resizebox{5in}{!}{\input{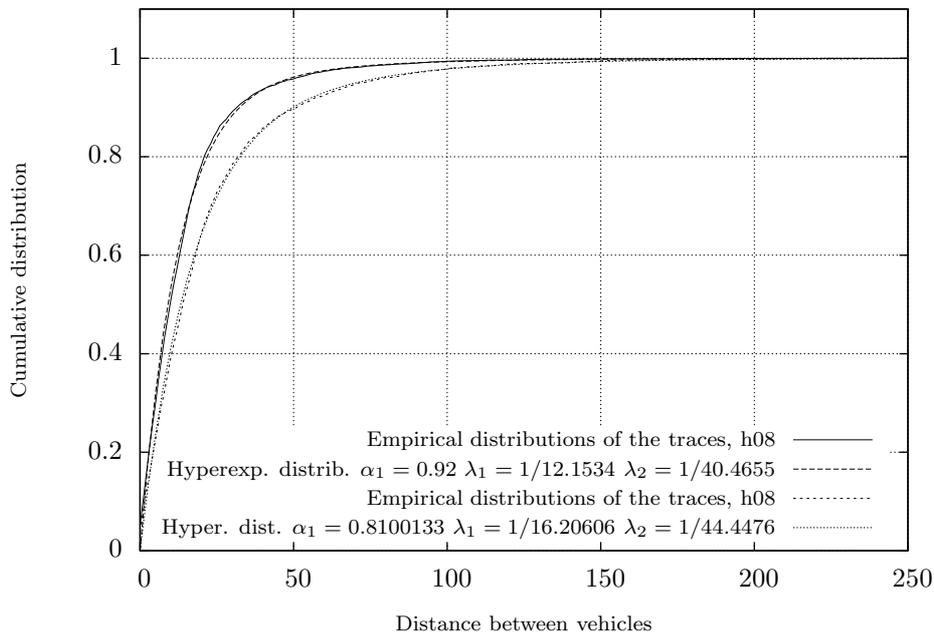}}
\caption{Cumulative distribution function of the distance (in meters) between the vehicles for the traces and for the corresponding hyperexponential distributions with fitted 
parameters}
\label{traceshyperexponentialinterdistances}
\end{figure}

On fig.~\ref{tracesexponentialinterdistances} the cumulative distributions of the distance in meters between the vehicles for the traces is compared to the
corresponding exponential distributions with same means. We observe that they differ. Then, on fig.~\ref{traceshyperexponentialinterdistances}, they are compared
to (\ref{F8h}) and (\ref{F11h}). The fitting is good.

\subsubsection{Comparison between the traces and the model}
\begin{figure}[!t]
\centering
\resizebox{5in}{!}{\input{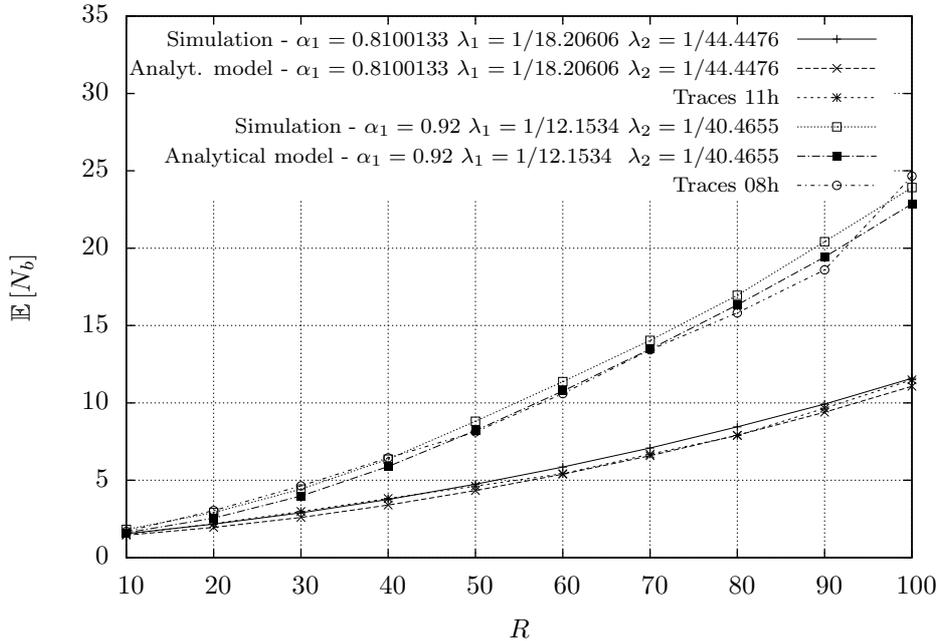}}
\caption{Average number of hops in a connected components, for the traces, the analytical model and the simulations with the corresponding hyperexponential distributions
(\ref{F8h}) and (\ref{F11h})}
\label{comparisontracessimulations}
\end{figure}

The comparison of the number of hops in the maximal shortest path (or number of hops in a connected component) for the traces, the analytical model and the simulations is 
given Fig.~\ref{comparisontracessimulations}. The match is not perfect due to the fact that the estimation of the distribution of the distances between the vehicles is also
no perfect. Actually, we decided to approximate it by hyperexponential laws as mixtures of two exponential distributions but maybe could it be more accurate to use a mixture
of more exponential laws than only two. Nevertheless, the results are sufficiently close to validate our model.

\subsubsection{Remark on the density of hop per unit length in a connected component}
\begin{figure}[!t]
\centering
\resizebox{5in}{!}{\input{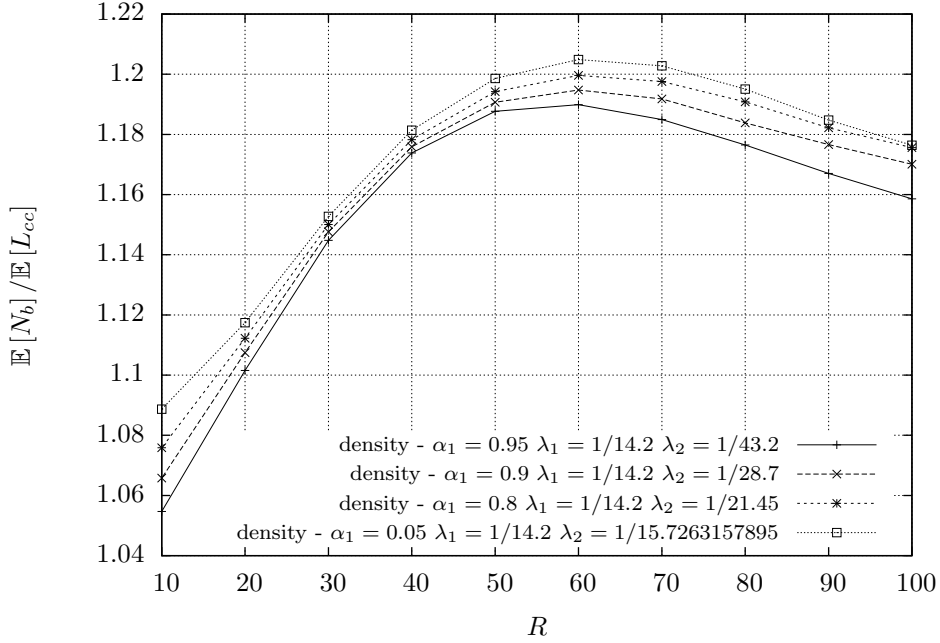}}
\caption{Average number of hops in a connected components divided by the average length of the connected component ("density of hops") in function of the coverage radius $R$, for the hyperexponential distributions
with various parameters $\alpha_1$, $\alpha_2=1-\alpha_1$, $\lambda_1$ and $\lambda_2$}
\label{densities}
\end{figure}

On Fig.~\ref{densities}, the average number of hops of the connected component divided by the length of the connected component is plotted in function 
of the coverage radius $R$ for the hyperexponential distributions with various parameters $\alpha_1$, $\alpha_2=1-\alpha_1$, $\lambda_1$ and $\lambda_2$.
The unit length is the radius $R$. To calculate the length of the connected component, we used (\ref{liushieq5}), which gives, in the case of the considered
hyperexponential process:
\begin{eqnarray}
\E{L_{cc}} & = & \frac{1}{\lambda_1\lambda_2}\frac{\alpha_1 \lambda_2\left(1-e^{-\lambda_1 R}\right)+\alpha_2\lambda_1\left(1-e^{-\lambda_2 R}\right)}{\alpha_1e^{-\lambda_1 R}
+\alpha_2 e^{-\lambda_2 R}}
\end{eqnarray}

All the curves corresponds to different parameters of the exponential mixture, but the resulting mean rate is kept constant.
Thus, it can be noticed that this density is not uniform with the traffic conditions: it depends on the traffic profile.

\section{Conclusion}
VANET modeling is a requirement to avoid complex simulations when designing new protocols. For studies involving routing mechanisms, the number of
hops along the shortest path must be characterized, especially (but not only) for VANET without infrastructure. In this paper, we propose a method to calculate the distribution
of the number of hops in the maximal shortest path of a vehicle connected component for any traffic assumption. Also, for the special case of the Poisson assumption,
we provide a closed-form expression of its z-transform for medium and dense networks. We derive also the average of this number and it is easy to obtain the other moments 
by following the same method as for deriving the expectation either analytically or numerically. 
We validated our approach with traces available at the Universitat Polit\`ecnica de Catalunya. 

These results can be exactly used to know the average number of hops for 
diffusion protocols in VANETs or other performance studies like in \citep{adel}. It can also be used to exactly estimate the
time needed to diffuse the information by multiplying the number of hops on the shortest path by the access time at each hop since it can be assumed an independance 
between the access to the medium and the spread of the connected component, even if both numbers increase with the road traffic intensity. It can also be used to approximate
the number of hops on a path over a certain distance separating the source and the destination by estimating the hop density of the shortest path and multiplying it by
the distance, the hop density being the average given in this paper divided by the average spread of the connected component given in \citep{altman}.

\appendix
\section{Demonstration of theorem \ref{interarrivees}}
\label{demonstrationthminterarrivees}
The probability that $\tau_n$ be less than $x_n$ knowing that $\tau_{n-1}=x_{n-1}$ is:
\begin{eqnarray}
\label{eqgenericinterar1}
F\left(\tau_n\leq x_n/\tau_{n-1}=x_{n-1}\right) & = & P\left(\tau_n\leq x_n/I_{n-1,1}\geq R-x_{n-1}\right)\nonumber\\
& = & \frac{P\left(\tau_n\leq x_n \cap I_{n-1,1}\geq R-x_{n-1}\right)}{P\left(I_{n-1,1}\geq R-x_{n-1}\right)}\nonumber\\
& = & \frac{P\left(\tau_n\leq x_n \cap I_{n-1,1}\geq R-x_{n-1}\right)}{1-F\left(R-x_{n-1}\right)}
\end{eqnarray}

and the probability to have $\tau_n\leq x_n \cap I_{n-1,1}$ is the probability to have any number of vehicles between $X_{n}+R-x_{n-1}$ and $x$ on one hand and on the other hand that
the next interdistance between the last vehicle just before $x$ and the next vehicle to be larger than $R-x_n$.
Its density is:

\begin{eqnarray}
\label{eqgenericinterar2}
\lefteqn{\frac{dP}{dx_n}\left(\tau_n=x_n\cap I_{n-1,1}\geq R-x_{n-1}\right)}\nonumber\\
& = & \frac{dP}{dx_n}\left(I_{n-1,1}=x_n\right)P\left(I_{n-1,2}\geq R-x_{n}\right)\nonumber\\
& + & \int_{i_{n-1,1}=R-x_{n-1}}^{x_n}\frac{dP\left(I_{n-1,1}=i_{n-1,1}\right)}{di_{n-1,1}}
\left(\frac{dP\left(I_{n-1,2}=i_{n-1,2}\right)}{di_{n-1,2}}\right)_{i_{n-1,2}=x_n-i_{n-1,1}}di_{n-1,1}P\left(I_{n-1,3}\geq R-x_{n}\right)\nonumber\\
& + & \int_{i_{n-1,1}=R-x_{n-1}}^{x_n}\int_{i_{n-1,2}=0}^{x_n-i_{n-1,1}}\frac{dP\left(I_{n-1,1}=i_{n-1,1}\right)}{di_{n-1,1}}\frac{dP\left(I_{n-1,2}=i_{n-1,2}\right)}{di_{n-1,2}}\times\nonumber\\
& & \left(\frac{dP\left(I_{n-1,3}=i_{n-1,3}\right)}{di_{n-1,3}}\right)_{i_{n-1,3}=x_n-i_{n-1,1}-i_{n-1,2}}di_{n-1,1}di_{n-1,2}P\left(I_{n-1,4}\geq R-x_{n}\right)\nonumber\\
& + & \hdots\nonumber\\
& = & f(x_n)\left[1-F(R-x_n)\right]\nonumber\\
& + & \int_{i_{n-1,1}=R-x_{n-1}}^{x_n}f(i_{n-1,1})f(x_n-i_{n-1,1})di_{n-1,1}\left[1-F(R-x_n)\right]\nonumber\\
& + & \int_{i_{n-1,1}=R-x_{n-1}}^{x_n}\int_{i_{n-1,2}=0}^{x_n-i_{n-1,1}}f(i_{n-1,1})f(i_{n-1,2})f(x_n-i_{n-1,1}-i_{n-1,2})di_{n-1,1}di_{n-1,2}\left[1-F(R-x_n)\right]\nonumber\\
& + & \hdots...
\end{eqnarray}
Let us define $g(x_n)$ such as:
\begin{eqnarray}
\label{eqgenericinterar2bis}
g(x_n) & = & f(x_n)\nonumber\\
& + & \int_{i_{n-1,1}=R-x_{n-1}}^{x_n}f(i_{n-1,1})f(x_n-i_{n-1,1})di_{n-1,1}\nonumber\\
& + & \int_{i_{n-1,1}=R-x_{n-1}}^{x_n}\int_{i_{n-1,2}=0}^{x_n-i_{n-1,1}}f(i_{n-1,1})f(i_{n-1,2})f(x_n-i_{n-1,1}-i_{n-1,2})di_{n-1,1}di_{n-1,2}\nonumber\\
& + & \hdots...
\end{eqnarray}
Then,
\begin{eqnarray}
\label{eqgenericinterar2ter}
\frac{dP}{dx_n}\left(\tau_n=x_n\cap I_{n-1,1}\geq R-x_{n-1}\right) & = & g(x_n)\left[1-F(R-x_n)\right]
\end{eqnarray}
and since for any function h 
\begin{eqnarray}
\label{eqgenericinterar3}
\int_{x=R-x_{n-1}}^{+\infty}e^{-sx}h(x)dx & = & e^{-s(R-x_{n-1})}{\cal L}\left[h(x+R-x_{n-1})\right],
\end{eqnarray}
we deduce from the definition of $g$ that
\begin{eqnarray}
\label{eqgenericinterar4}
\int_{x=R-x_{n-1}}^{+\infty}e^{-sx}g(x)dx & = & e^{-s(R-x_{n-1})}{\cal L}\left[g(x+R-x_{n-1})\right] \nonumber\\
& = & e^{-s(R-x_{n-1})}{\cal L}\left[f(x+R-x_{n-1})\right]\displaystyle\sum_{k=0}^{+\infty}{\cal L}\left[f(x)\right]^k\nonumber
\end{eqnarray}
so that
\begin{eqnarray}
\label{eqgenericinterar5}
{\cal L}\left[g(x+R-x_{n-1})\right] & = & \frac{{\cal L}\left[f(x+R-x_{n-1})\right]}{1-{\cal L}\left[f(x))\right]}
\end{eqnarray}
Combining (\ref{eqgenericinterar2ter}) and (\ref{eqgenericinterar5}), integrating on $x_n$ between $R-x_{n-1}$ and $x$ and using (\ref{eqgenericinterar1}), we get
(\ref{eqgenericinterdist1}).

\section{Demonstration of theorem \ref{casspecial}}
\label{demonstrationGrandtheoremeCasSpecial}
The idea of the proof is to begin by finding the expression of $M_1(z)$, given by lemma~\ref{grandlemme}, which is proved itself 
in \ref{demonstrationgrandlemme}. Lemmas  \ref{lemma3}, \ref{lemma4}, \ref{lemma5} and \ref{lemma6} are then used to simplify
the expression of $M_1(z)$ given by lemma~\ref{grandlemme}. The lemma \ref{lemmeconvergence} gives the condition of convergence of
$M_1(z)$.

Let us define the following notations.
\begin{eqnarray}
\label{les_bi}b_i&=&\rho'^i\left(\sum_{j=0}^i(-1)^j\frac{\lambda'^j}{j!}C_{2i-j}^i\right)\\
\label{les_ci}c_i&=&\rho'^iC_{2i-1}^i\\
\label{a2ki}a_{2k,i}&=&\rho'^{k-1}(-1)^iC_{2k-i-3}^{k-2}\\
\label{a2kplus1i}a_{2k+1,i}&=&\rho'^{k}(-1)^iC_{2k-i-1}^{k-1}.
\end{eqnarray}

Let us also define the following z-transforms.
\begin{eqnarray}
\forall i>0,A_i^{(e)}(z)& = &\sum_{k=i+1}^{+\infty}a_{2k,i}z^{2k}\\
A_0^{(e)}(z) & = & \sum_{k={\bf 2}}^{+\infty}a_{2k,0}z^{2k}\\
\label{Aio}\forall i>0,A_i^{(o)}(z) & = & \sum_{k=i}^{+\infty}a_{2k+1,i}z^{2k+1}\\
A_0^{(o)}(z) & = & \sum_{k={\bf 1}}^{+\infty}a_{2k+1,0}z^{2k+1}
\end{eqnarray}

\begin{lemma}
\label{lemmeconvergence}
The series $\displaystyle\sum^{+\infty}b_iz^{2i+1}$ converges for $\lambda'>\ln4+\ln z^2$.
\end{lemma}

\begin{pf}
\begin{eqnarray}
b_i&=&\rho'^i\left(\sum_{j=0}^i(-1)^j\frac{\lambda'^j}{j!}C_{2i-j}^i\right)\nonumber\\
& = & \rho'^i\left(C_{2i}^i\sum_{j=0}^i\frac{(-\lambda')^j}{j!}\frac{C_{2i-j}^i}{C_{2i}^i}\right)\nonumber\\
& = & \rho'^i\left(C_{2i}^i\sum_{j=0}^i\frac{(-\lambda')^j}{j!}\frac{(2i-j)!i!}{(2i)!(i-j)!}\right)
\end{eqnarray}
But 
$$\sum_{j=0}^i\frac{(-\lambda')^j}{j!}\frac{(2i-j)!i!}{(2i)!(i-j)!}$$ 
is bounded because 
\begin{eqnarray}
\frac{(2i-j)!i!}{(2i)!(i-j)!} & = & \frac{i(i-1)\cdots (i-j+1)}{2i(2i-1)\cdots (2i-j+1)}\nonumber\\
& < & 1.
\end{eqnarray}
Moreover, in the neighborhood of $+\infty$, by denoting $Cst$ a constant term, and using Stirling's formula,
\begin{eqnarray}
b_iz^{2i+1} & \sim & \rho'^i Cst \times C_{2i}^iz^{2i+1}\nonumber\\
 & \sim & \rho'^i Cst \frac{(2i)!}{i!i!}z^{2i+1}\nonumber\\
 & \sim & \rho'^i Cst \frac{\sqrt{2\pi 2 i}\left(\frac{2i}{e}\right)^{2i}}{\left[\sqrt{2\pi i}\left(\frac{i}{e}\right)^i\right]^2}z^{2i+1}\nonumber\\
 & \sim & Cst \frac{e^{i\ln 4-i\lambda'}}{\sqrt{\pi i}}z^{2i+1}
\end{eqnarray}
Then,  $\displaystyle\sum^{+\infty}b_iz^{2i+1}$ converges if and only if $\lambda'>\ln4+2\ln z$.
\end{pf}

Similarly, it can be shown that the other sums converges only when $\lambda'>\ln4$ for $z$ in the neighborhood of 1.
Then, for $\lambda'>\ln4$ for $z$ in the neighborhood of 1, we have the following lemma.

\begin{lemma}
\label{grandlemme}
\begin{eqnarray}
\label{M1dezDuGrandThereome}
M_1(z)
 &=&\frac{1}{\rho z^2\left[1-b_0z+\displaystyle\sum_{i=1}^{+\infty}\left(c_iz^{2i}-b_iz^{2i+1}\right)\right]}\times\\
& & \left[(z-1)\sum_{i=0}^{+\infty}\frac{\lambda'^i}{i!}\left(\mathfrak{M}_{i,1}A_i^{(o)}(z)+\mathfrak{M}_{i,2}A_i^{(e)}(z)\right)\right.\nonumber\\
& & \left.+\left(\mathfrak{M}_{0,1}z+\mathfrak{M}_{0,2}z^2-\mathfrak{M}_{0,1}z^2\right)\sum_{i=1}^{+\infty}c_iz^{2i}
-\mathfrak{M}_{0,2}z^2\sum_{i=1}^{+\infty}b_iz^{2i+1}\right]\nonumber
\end{eqnarray}
where, for all $\alpha\geq0$,
\begin{eqnarray}
\label{equationMalpha1}
\mathfrak{M}_{\alpha,1} & = & \sum_{i=0}^\alpha \frac{(-1)^i\alpha!}{(\alpha-i)!}\frac{1}{\lambda'^i}+(-1)^{\alpha+1}\frac{\alpha!}{\lambda'^\alpha}\rho'
\end{eqnarray}
and, for all $\alpha\geq0$,
\begin{eqnarray}
\label{equationMalpha2}
\mathfrak{M}_{\alpha,2} & = & \sum_{i=0}^\alpha \frac{(-1)^i\alpha!}{(\alpha-i)!}\frac{1}{\lambda'^i}-\frac{\rho}{\alpha+1}-(-1)^\alpha\frac{\alpha!}{\lambda'}\rho'
\end{eqnarray}
\end{lemma}

\begin{pf}
See Appendix~\ref{demonstrationgrandlemme}.
\end{pf}

At last, by combining lemmas \ref{lemma3}, \ref{lemma4}, \ref{lemma5} and \ref{lemma6}, this expression of $M_1(z)$ can be simplified into (\ref{M1dezCasSpecial}).

Let
\begin{eqnarray}
\label{ualphak}
u_{\alpha,k} & = & \frac{\lambda'^\alpha}{\alpha!}\mathfrak{M}_{\alpha,k}.
\end{eqnarray}

\begin{lemma}
\label{lemma3}
If $\lambda R\geq \ln 4$, 
\begin{eqnarray}
\label{premiersommeuA}\sum_{\alpha=1}^{+\infty} u_{\alpha,1}A_\alpha^{(o)}(z)
& = & \frac{f(z)}{1+\frac{1}{2}g(z)}\left[\frac{1}{2}g(z)(\rho'-1)+e^{\frac{1}{2}\lambda'g(z)}-1\right]
\end{eqnarray}
where
\begin{eqnarray}
f(z) & = & \frac{2z(\rho'z^2)}{\sqrt{1-4\rho' z^2}-(1-4\rho z^2)}
\end{eqnarray}
and
\begin{eqnarray}
g(z) & = & \sqrt{1-4\rho' z^2} -1
\end{eqnarray}
\end{lemma}

\begin{pf}
From equations (\ref{a2kplus1i}) and (\ref{Aio}), $\forall i>0,$
\begin{eqnarray}
A_i^{(o)}(z) & = & \rho'^i(-1)^iz^{2i+1}\sum_{k=0}^{+\infty}\rho'^kC_{2k+i-1}^kz^{2k}.
\end{eqnarray}
But,
\begin{eqnarray}
\frac{\rho'^{k+1}C_{2(k+1)+i-1}^{k+1}}{\rho'^kC_{2k+i-1}^k} &=& 4\rho'\frac{\left[k+\frac{i+1}{2}\right]\left[k+\frac{i}{2}\right]}{(k+1)(k+i)}
\end{eqnarray}
so that the series $\displaystyle\sum_{k=0}^{+\infty}\rho'^kC_{2k+i-1}^kz^{2k}$ is the hypergeometric series: 
\begin{eqnarray}
\displaystyle\sum_{k=0}^{+\infty}\rho'^kC_{2k+i-1}^kz^{2k}
&=& _2F_1\left(\frac{i+1}{2},\frac{i}{2},i,4\rho'z^2\right).
\end{eqnarray}
Also, 
\begin{eqnarray}
_2F_1\left(a,b,2b,\frac{4u}{(u+1)^2}\right)
 & = & (u+1)^{2a}{_2F_1}\left(a,a-b+\frac{1}{2},b+\frac{1}{2},u^2\right)\quad
\end{eqnarray}
and
\begin{eqnarray}
_2F_1(a,1,a,u^2) & = & (1-u^2)^{-1}.
\end{eqnarray}
Thus, by chosing $u$ so that 
\begin{eqnarray}
\label{conditionexistenceu}\displaystyle\frac{4u}{(u+1)^2}=4\rho'z^2,
\end{eqnarray}
we have
\begin{eqnarray}
\forall i>0,
A_i^{(o)}(z) & = & f(z)\frac{1}{2^i}g(z)^i.
\end{eqnarray}
At last,
\begin{eqnarray}
\sum_{\alpha=1}^{+\infty}u_{\alpha,1}A_\alpha^{(o)}(z)
& = & \sum_{\alpha=1}^{+\infty}(-1)^\alpha\left[\sum_{j=0}^\alpha(-1)^j\frac{\lambda'^j}{j!}-\rho'\right]f(z)\frac{1}{2^\alpha}g(z)^\alpha\quad
\end{eqnarray}
and noting that
\begin{eqnarray}
\label{identitedoublesommes}\sum_{\alpha=1}^{+\infty}\sum_{j=0}^\alpha v_jw_\alpha = \sum_{\alpha=1}^{+\infty}v_0w_\alpha+\sum_{j=1}^{+\infty}\sum_{\alpha=j}^{+\infty}v_jw_\alpha,
\end{eqnarray}
where $(v_n)$ and $(w_n)$ are two arbitrary sequences, we get (\ref{premiersommeuA}).
Note that it exists a $u$ solution of (\ref{conditionexistenceu}) only if $\lambda R\geq\ln 4$.
\end{pf}

\vspace{1ex}
Similarly, by noting that $a_{2k,i}=a_{2k-1,i}$, the following lemma can be proved.
\begin{lemma}
\label{lemma4}
If $\lambda R\geq \ln 4$,
\begin{eqnarray}
\label{deuxiemesommeuA}\sum_{\alpha=1}^{+\infty} u_{\alpha,2}A_\alpha^{(e)}(z)
& = & \frac{2\rho'zf(z)}{g(z)}\left[1+\frac{1}{2}\lambda'g(z)-e^{\frac{1}{2}\lambda'g(z)}\right]
 +\frac{1}{2}(\rho'-1)\frac{zf(z)g(z)}{1+\frac{1}{2}g(z)}+zf(z)\frac{e^{\frac{1}{2}\lambda'g(z)}-1}{1+\frac{1}{2}g(z)}\qquad\qquad
\end{eqnarray}
\end{lemma}

\begin{lemma}
\label{lemma5}
If $\lambda R\geq \ln 4$,
\begin{eqnarray}
\label{sommeciz2i}\sum_{i=1}^{+\infty} c_iz^{2i}& = & \frac{1}{4\rho'z^3}f(z)g(z)^2.
\end{eqnarray}
\end{lemma}

\begin{pf}
On one hand,
\begin{eqnarray}
\sum_{i=1}^{+\infty} c_iz^{2i}& = &\rho'z^2{_2F_1\left(\frac{3}{2},1,2,4\rho'z^2\right)}
\end{eqnarray}
and on the other hand,
\begin{eqnarray}
_2F_1\left(\frac{3}{2},1,2,\frac{4u}{(1+u)^2}\right)
 & = & (u+1)^3{_2F_1\left(\frac{3}{2},1,\frac{3}{2},u^2\right)}\\
 & = & \frac{(u+1)^3}{1-u^2}.\nonumber
\end{eqnarray}
Thus, with $u$ solution of (\ref{conditionexistenceu}), (\ref{sommeciz2i}) is proved.
\end{pf}

\begin{lemma}
\label{lemma6}
If $\lambda R\geq \ln 4$,
\begin{eqnarray}
\label{sommebiz2iplus1}\sum_{i=1}^{+\infty} b_iz^{2i+1}& = & \frac{z\left[e^{\frac{1}{2}\lambda'g(z)}-g(z)-1\right]}{1+g(z)}.
\end{eqnarray}
\end{lemma}

\begin{pf}
This lemma can be proved by using identity (\ref{identitedoublesommes}) and the following ones:
\begin{eqnarray}
_2F_1\left(\frac{1}{2},1,1,4\rho'z^2\right) & = & \frac{1}{\sqrt{1-4\rho'z^2}}
\end{eqnarray}
and
\begin{eqnarray}
_2F_1\left(a,b,2b,\frac{4u}{(u+1)^2}\right)
 & = & (1+u)^{2a}{_2F_1\left(a,a-b+\frac{1}{2},b+\frac{1}{2},u^2\right)}\quad
\end{eqnarray}

\end{pf}

\section{Demonstration of lemma\ref{grandlemme}}
\label{demonstrationgrandlemme}
In this section, we derive $M_1(z)$.
To find the elements of the sequence $\left(\mathfrak{M}_{1,k}\right)_k$ it is necessary to solve the double recurrence, both in $\alpha$ and $k$. 
That is why we will proceed in two steps.
The set of the positive values of $\alpha$ and $k$ constitutes a quarter plane. 
In the first step, we will bring back the recurrence in the borders of the quarter plane by eliminating the terms in $\mathfrak{M}_{\alpha,k}$ for which
$\alpha>0$ and $k>2$ and keep only the terms of the form $\mathfrak{M}_{0,k}$ for $k>2$ and the terms of the form
 $\mathfrak{M}_{\alpha,1}$ or $\mathfrak{M}_{\alpha,2}$. 
Actually, $\mathfrak{M}_{\alpha,1}$ and $\mathfrak{M}_{\alpha,2}$ are easy to calculate.
This way, we will obtain a new recurrence expressing $\mathfrak{M}_{0,k}$ in function of $\mathfrak{M}_{0,i}$ where $i<k$, $\mathfrak{M}_{\alpha,1}$ 
and $\mathfrak{M}_{\alpha,2}$. In the second step, we will solve this last recurrence in order to find $\mathfrak{M}_{0,k}$ in function of 
the $\mathfrak{M}_{\alpha,1}$ and $\mathfrak{M}_{\alpha,2}$. Then, using equation (\ref{M0k}), it is easy to deduce $\mathfrak{M}_{1,k}$.

\subsection{First step - expressing $\mathfrak{M}_{0,k}$ in function of $\mathfrak{M}_{0,i}, i<k$, $\mathfrak{M}_{\alpha,1}$ and $\mathfrak{M}_{\alpha,2}$}

By injecting (\ref{ualphak}) in (\ref{M0k}) and (\ref{Malphak}), a simpler system is obtained:
\begin{numcases}{}
 \label{sysu0k}u_{0,k} =u_{0,k-1} -\rho' u_{1,k-2}\\
 \label{sysualphak}u_{\alpha,k} = \frac{\lambda'^\alpha}{\alpha!}u_{0,k-1}-u_{\alpha-1,k}-
\rho'u_{\alpha+1,k-2}
\end{numcases}

Let $V_k(z)$ be the z-transform of $u_{\alpha,k}$ relatively to $\alpha$:

\begin{eqnarray}
V_k(z) & = & \sum_{\alpha=0}^{+\infty}u_{\alpha,k}z^\alpha
\end{eqnarray}

Multiplying (\ref{sysualphak}) by $z^\alpha$ and summing over $\alpha\geq 1$, 
\begin{eqnarray}
V_k(z) 
& = & \frac{e^{\lambda'z}}{1+z}u_{0,k-1}+\frac{\rho'}{z(1+z)}u_{0,k-2}-\frac{\rho'}{z(1+z)}V_{k-2}(z)
\end{eqnarray}

By recurrence, 
\begin{eqnarray}
\label{V2k}
\lefteqn{V_{2k}(z)}\\
& = &\left[-\frac{\rho'}{z(1+z)}\right]^{k-1}V_2(z)
-\sum_{i=1}^{k-1}\left[-\frac{\rho'}{z(1+z)}\right]^iu_{0,2(k-i)}
+\frac{e^{\lambda'z}}{1+z}\sum_{i=0}^{k-2}\left(-\frac{\rho'}{z(1+z)}\right)^iu_{0,2(k-i)-1}\nonumber
\end{eqnarray}
\begin{eqnarray}
\label{V2kplus1}
\lefteqn{V_{2k+1}(z)}\\
&= &\left[-\frac{\rho'}{z(1+z)}\right]^{k-1}V_1(z)
-\sum_{i=1}^{k}\left[-\frac{\rho'}{z(1+z)}\right]^iu_{0,2(k-i)+1}
+\frac{e^{\lambda'z}}{1+z}\sum_{i=0}^{k-1}\left(-\frac{\rho'}{z(1+z)}\right)^iu_{0,2(k-i)}.\nonumber
\end{eqnarray}

In order to find the elements of the type $u_{0,k}$, the 0 degree in $z$ of these equations should be extracted.
Knowing that 
\begin{eqnarray}
\forall z\in]-1;+1[, 
(1+z)^{-\alpha} & = & 1+\sum_{n=1}^{+\infty}(-1)^nC_{\alpha+n-1}^nz^n,
\end{eqnarray}
equating the 0 degrees in $z$ of both sides of equation (\ref{V2k}) gives
\begin{eqnarray}
\label{recintermediaire1}\lefteqn{u_{0,2k} }\\
&=& \rho'^{k-1}\sum_{i=0}^{k-1}(-1)^iC_{2k-i-3}^{k-2}u_{i,2}-\sum_{i=1}^{k-1}\rho'^iC_{2i-1}^iu_{0,2(k-i)}
 +\sum_{i=0}^{k-2}\rho'^i\left(\sum_{j=0}^i(-1)^j\frac{\lambda'^j}{j!}C_{2i-j}^i\right)u_{0,2(k-i)-1}\nonumber
\end{eqnarray}

and the same for equation (\ref{V2kplus1}):
\begin{eqnarray}
\label{recintermediaire2}\lefteqn{u_{0,2k+1} }\\
&=& \rho'^{k}\sum_{i=0}^{k}(-1)^iC_{2k-i-1}^{k-1}u_{i,1}-\sum_{i=1}^{k}\rho'^iC_{2i-1}^iu_{0,2(k-i)+1}
+\sum_{i=0}^{k-1}\rho'^i\left(\sum_{j=0}^i(-1)^j\frac{\lambda'^j}{j!}C_{2i-j}^i\right)u_{0,2(k-i)}.\nonumber
\end{eqnarray}

Now, the system of equations (\ref{recintermediaire1}) and (\ref{recintermediaire2}) gives an expression of $u_{0,k}$ in function of $u_{0,i}$ for $i<k$. 
In the next section, we solve this recurrence by searching the z-transform $U_0(z)$ of $u_{0,k}$ and making the link between $U_0(z)$ and $M_1(z)$.

\subsection{Second step - finding $M_1(z)$}

Let us define the following z-transforms.
\begin{eqnarray}
U_0^{(e)}(z) & = & \sum_{k=1}^{+\infty}u_{0,2k}z^{2k}\\
U_0^{(o)}(z) & = & \sum_{k=0}^{+\infty}u_{0,2k+1}z^{2k+1}\\
U_0(z) & = & U_0^{(e)}(z)+U_0^{(o)}(z)\\
\forall i\geq 0, U_i(z) & = & \sum_{k=0}^{+\infty}u_{i,k}z^k
\end{eqnarray}

Then, using notations (\ref{les_bi}), (\ref{les_ci}), (\ref{a2ki}) and (\ref{a2kplus1i}), the system of equations (\ref{recintermediaire1}) and (\ref{recintermediaire2}) becomes
\begin{numcases}{}
\label{equationu02k}u_{0,2k}=
\quad\sum_{i=0}^{k-1}a_{2k,i}u_{i,2}+\sum_{i=0}^{k-2}b_iu_{0,2(k-i)-1}-\sum_{i=1}^{k-1}c_iu_{0,2(k-i)}\\
\label{equationu02kplus1}u_{0,2k+1}=
\quad\sum_{i=0}^ka_{2k+1,i}u_{i,1}+\sum_{i=0}^{k-1}b_iu_{0,2(k-i)}-\sum_{i=1}^kc_iu_{0,2(k-i)+1}
\end{numcases}

Multiplying equation (\ref{equationu02k}) by $z^{2k}$ and summing for $k\geq2$, multiplying equation (\ref{equationu02kplus1}) by $z^{2k+1}$ and
summing for $k\geq1$ and summing both leads to
\begin{eqnarray}
\lefteqn{\sum_{k\geq3}^{+\infty} u_{0,k}z^k=}\\
 && \sum_{i=0}^{+\infty}\left[u_{i,1}A_i^{(o)}(z)+u_{i,2}A_i^{(e)}(z)\right]-u_{0,1}z\sum_{i=0}^{+\infty}b_iz^{2i+1}
+\left[U_0^{(e)}(z)+U_0^{(o)}(z)\right]\left[\sum_{i=0}^{+\infty}b_iz^{2i+1}-\sum_{i=1}^{+\infty}c_iz^{2i}\right]\nonumber
\end{eqnarray}

or, equivalently,
\begin{eqnarray}
\label{equationU0}\lefteqn{U_0(z)\left[1-b_0z+\sum_{i=1}^{+\infty}\left(c_iz^{2i}-b_iz^{2i+1}\right)\right]}\\
&=&\sum_{i=0}^{+\infty}\left[u_{i,1}A_i^{(o)}(z)+u_{i,2}A_i^{(e)}(z)\right]
+u_{0,1}z+u_{0,2}z^2-u_{0,1}z\sum_{i=0}^{+\infty}b_iz^{2i+1}\nonumber
\end{eqnarray}

Moreover, multiplying equations (\ref{sysu0k}) by $z^k$ for every $k$ and summing for all $k\geq3$, leads to
\begin{eqnarray}
\label{equationU0U1}
U_0(z)-u_{0,2}z^2-u_{0,1}z
&=&z\left[U_0(z)-u_{0,1}z\right]-\rho'z^2U_1(z)
\end{eqnarray}

and, since $\mathfrak{M}_{1,k}=u_{1,k}\times\frac{1!}{\lambda'^1}$,
\begin{eqnarray}
\label{equationM1U1}
M_1(z) & = & \frac{1}{\lambda'}U_1(z).
\end{eqnarray}

Combining (\ref{equationU0U1}) and (\ref{equationM1U1}),
\begin{eqnarray}
M_1(z)&=&\frac{u_{0,2}-u_{0,1}}{\rho}+\frac{u_{0,1}}{\rho}\frac{1}{z}+\frac{z-1}{\rho z^2}U_0(z)
\end{eqnarray}

Finally, thanks to (\ref{equationU0}),
\begin{eqnarray}
\label{equationM1}
M_1(z)
 &=&\frac{1}{\rho z^2\left[1-b_0z+\displaystyle\sum_{i=1}^{+\infty}\left(c_iz^{2i}-b_iz^{2i+1}\right)\right]}\times\\
& & \left[(z-1)\sum_{i=0}^{+\infty}\left(u_{i,1}A_i^{(o)}(z)+u_{i,2}A_i^{(e)}(z)\right)\right.\nonumber\\
& & \left.+\left(u_{0,1}z+u_{0,2}z^2-u_{0,1}z^2\right)\sum_{i=1}^{+\infty}c_iz^{2i}
-u_{0,2}z^2\sum_{i=1}^{+\infty}b_iz^{2i+1}\right]\nonumber
\end{eqnarray}

Replacing $u_{\alpha,k}$ by $\mathfrak{M}_{\alpha,k}$, we get (\ref{M1dezDuGrandThereome}).\\

At last, by integrating by parts, $\mathfrak{M}_{\alpha,1}=1-\frac{\alpha}{\lambda'}\mathfrak{M}_{\alpha-1,1}$ and 
$\mathfrak{M}_{\alpha,2}=1-\rho'-\frac{\rho}{\alpha+1}-\frac{\alpha}{\lambda'}\mathfrak{M}_{\alpha-1,2}$.
Then, by recurrence (\ref{equationMalpha1}) and (\ref{equationMalpha2}) are obtained.

\vspace{0.5cm}
\begin{wrapfigure}{l}{0.1\textwidth}
\vspace{-24pt}
  \begin{center}
    \includegraphics[width=2cm]{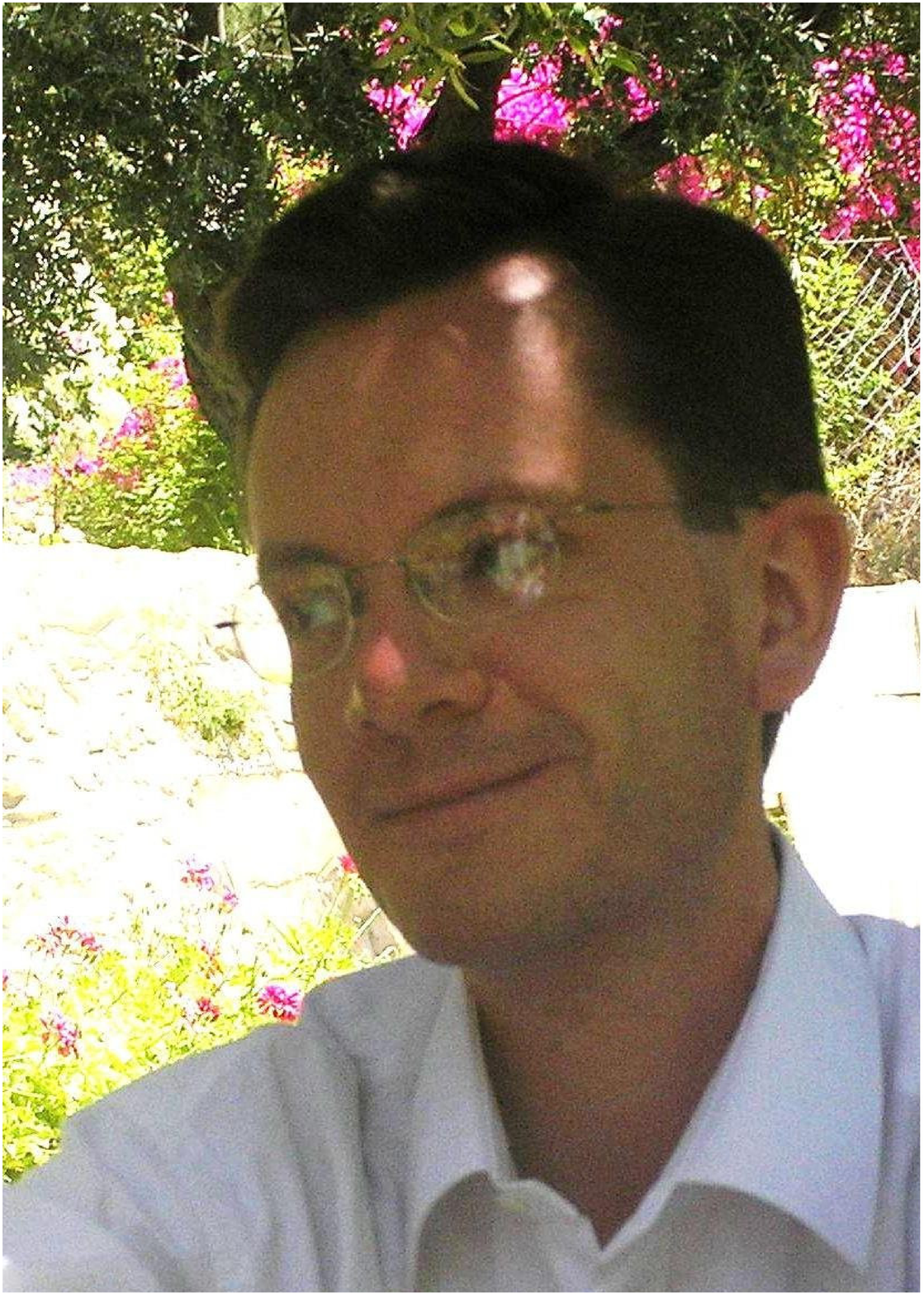}
  \end{center}
\vspace{-20pt}
\end{wrapfigure}
\noindent
{\bf Michel Marot} was born in Paris in 1973. He received the Ph.D. and habilitation degrees in computer networks from University of Paris 6 in 2001 and 2010. 
He is a professor in the Telecommunication Networks and services department at the Institut Mines-T\'el\'ecom, T\'el\'ecom SudParis, France.
His current research interests are mainly on network performance and self-organization in wireless networks, ad-hoc and sensor 
networks, vehicular networks, context aware adaptation. His other research interests include mobility modeling, complex systems, and queuing theory. \\

\begin{wrapfigure}{l}{0.1\textwidth}
\vspace{-24pt}
  \begin{center}
    \includegraphics[width=2cm]{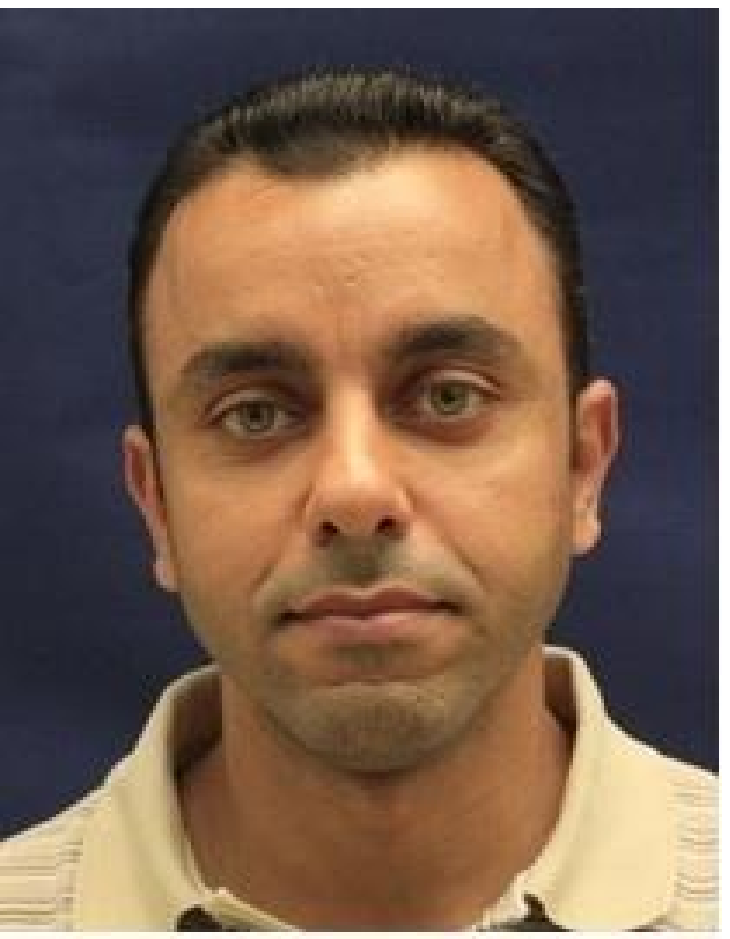}
  \end{center}
\vspace{-20pt}
\end{wrapfigure}
\noindent
{\bf Adel Mounir Said} was born in Cairo in 1978. He received the B.S. from the Higher Technological Institute in 2001, and M.S. in wireless communication 
from the Arab Academy for science, Technology \& Maritime Transport in 2007. He got the Egyptian Syndicate of Engineering award in 2001 for his scholar achievement. 
He is PhD student in l'Universit\'e Pierre et Marie Curie conjointement \`a T\'el\'ecom SudParis and Ain Shams University. Adel Mounir Said is currently assistant 
lecturer in the National Telecommunication Institute, Egypt. He is specialist in fixed mobile convergence, wireless sensor, VoIP, NGN, IMS, and vehicular networks.\\

\newpage
\begin{wrapfigure}{l}{0.1\textwidth}
\vspace{-24pt}
  \begin{center}
    \includegraphics[width=2cm]{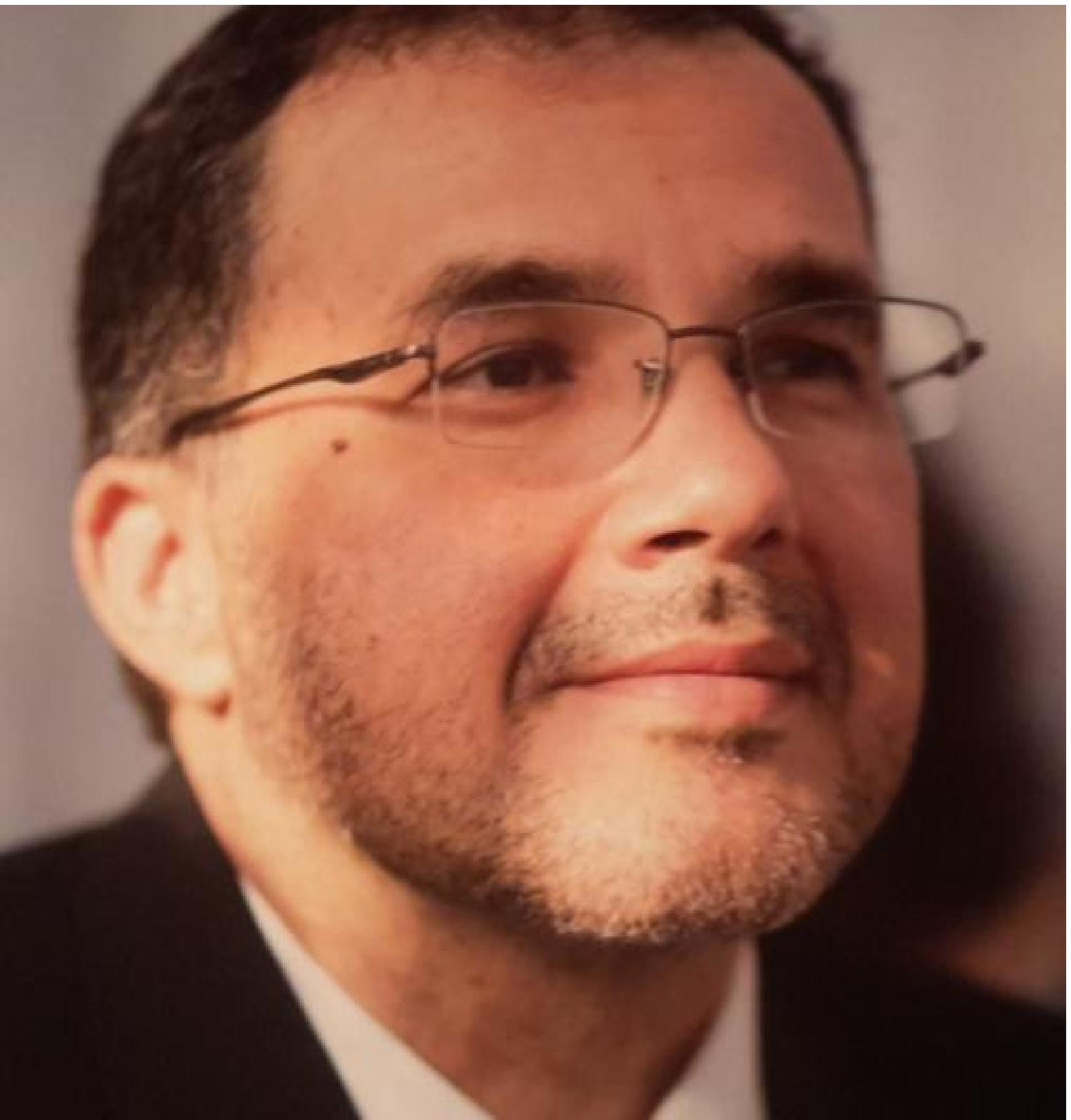}
  \end{center}
\vspace{-20pt}
\end{wrapfigure}
\noindent
{\bf Hossam Afifi} is professor in computer networks and network security. He works at T\'el\'ecom SudParis in Saclay France.  
His research interests cover wireless network design and performance evaluation and networking security protocols. 
He is active in IEEE WLAN standardization and IETF protocol specification. He obtained a Ph.D. from INRIA Sophia Antipolis (France). 
He has several sabbatical stays in the USA, especially in Washington University St. Louis and Nokia Research Moutain View (CA). 
He was appointed as associate professor at T\'el\'ecom Bretagne for seven years and now as professor in the same Institut Mines T\'el\'ecom 
structure with T\'el\'ecom SudParis.


\end{document}